# Dependent Dirichlet Priors and Optimal Linear Estimators for Belief Net Parameters

**Peter M. Hooper**
Dept. of Mathematical & Statistical Sciences
University of Alberta
Edmonton, AB T6G 2G1 Canada

## Abstract

A Bayesian belief network is a model of a joint distribution over a finite set of variables, with a DAG structure representing immediate dependencies among the variables. For each node, a table of parameters (CP-table) represents local conditional probabilities, with rows indexed by conditioning events (assignments to parents). CP-table rows are usually modeled as independent random vectors, each assigned a Dirichlet prior distribution. The assumption that rows are independent permits a relatively simple analysis but may not reflect actual prior opinion about the parameters. Rows representing similar conditioning events often have similar conditional probabilities. This paper introduces a more flexible family of "dependent Dirichlet" prior distributions, where rows are not necessarily independent. Simple methods are developed to approximate the Bayes estimators of CP-table parameters with optimal linear estimators; i.e., linear combinations of sample proportions and prior means. This approach yields more efficient estimators by sharing information among rows. Improvements in efficiency can be substantial when a CP-table has many rows and samples sizes are small.

## 1 INTRODUCTION

Bayesian belief nets provide concise models of joint probability distributions used in a wide variety of applications. Belief nets are typically constructed by first finding an appropriate structure (by interviewing an expert or by selecting a good model from training data), then using a training sample to estimate the parameters (Heckerman, 1999). Bayesian methods are usually employed in this second step, with the prior distribution satisfying a local independence assumption (Spiegelhalter and Lauritzen, 1990); i.e., CP-tables are assumed to have independent Dirichlet rows. Golinelli, Madigan, and Consonni (1999) questioned the appropriateness of this assumption and developed estimators based on a more flexible family of priors called hierarchical partition models. The present paper follows a broadly similar approach, developing estimators based on a new family of priors. These new estimators exploit similarities among conditioning events to share information among CP-table rows.

To introduce ideas, suppose we have a child node $X$ with two parents $A$ and $B$. Suppose all three variables are binary (0/1). The CP-table entries for node $X$ consist of conditional probabilities

$$\theta_{x|ab} = Pr\{X = x \,|\, A = a, B = b, \mathbf{\Theta}\}.$$

The symbol $\mathbf{\Theta}$ represents the vector of all CP-table entries in the belief net. We include $\mathbf{\Theta}$ in the conditioning event to reflect the assumption that the parameters $\theta_{x|ab}$ are random variables with some joint prior distribution. The mean of the prior distribution is $E(\theta_{x|ab}) = Pr\{X = x \,|\, A = a, B = b\}$. The CP-table has 4 rows indexed by $\langle a, b\rangle$ and 2 columns indexed by $x$. The probabilities in each row sum to one, so in the binary case it suffices to consider a single column; say $x = 1$.

Suppose our training data consists of a random sample of $n$ complete tuples. Let $m_{xab}$ denote the number of tuples with $\langle X, A, B\rangle = \langle x, a, b\rangle$, and put $n_{ab} = \sum_x m_{xab}$. If $n_{ab} > 0$, then we can define the sample proportion $p_{x|ab} = m_{xab}/n_{ab}$. Sample proportions are unbiased estimators, but have high variance when $n_{ab}$ is small and are undefined when $n_{ab} = 0$. Bayesian estimators reduce variance by allowing bias. The usual priors assume that the CP-table rows are independent Dirichlet random vectors. For the binary variable $X$, this is equivalent to assuming that $\{\theta_{1|ab}\}$ are inde-pendent Beta variables. The hyperparameters defining the Dirichlet/Beta distributions can be selected to reflect expert opinion by specifying the prior means and variances of the CP-table entries. In the absence of expert opinion, the Beta(1,1) distribution is commonly selected since this has a uniform density over the unit interval $[0, 1]$.

Priors with independent Dirichlet rows are popular in large part because they form a conjugate family of distributions; i.e., the posterior distribution (the conditional distribution of the CP-table parameters given the data) also has independent Dirichlet rows, but with hyperparameters updated to reflect information from the data. This property makes Bayesian estimates easy to calculate and interpret. The mean posterior (MP) estimator is the mean of the posterior distribution. For uniform priors, the MP estimator is obtained by adding 1 to each count $m_{xab}$ before calculating sample proportions. More generally, MP estimators are weighted averages of the sample proportion $p_{x|ab}$ and the prior mean $E(\theta_{x|ab})$. When $n_{ab} = 0$, all of the weight is on the prior mean. When $n_{ab}$ is large, most of the weight is on $p_{x|ab}$.

While simple to use, priors with independent rows may fail to reflect actual prior opinion. The independence assumption implies that knowledge of some rows would provide no information about the other rows. In my view, it seems more reasonable to expect CP-table entries in the same column but different rows to be positively correlated, with stronger correlations for more similar conditioning events. E.g., in the absence of information to the contrary, I would expect $\theta_{x|00}$ to be closer to $\theta_{x|01}$ than to $\theta_{x|11}$. Correlations need not be symmetric. I might expect $\theta_{x|00}$ to be closer to $\theta_{x|01}$ than to $\theta_{x|10}$ if I believe that $X$ is more strongly related to $A$ than to $B$. These considerations suggest that an estimator of $\theta_{x|00}$ might usefully incorporate information from $p_{x|01}$, $p_{x|10}$, and perhaps $p_{x|11}$, in addition to $p_{x|00}$ and the prior mean $E(\theta_{x|00})$.

This paper develops a theoretical framework to implement these ideas. I introduce a flexible family of dependent Dirichlet (DD) prior distributions. Given a DD prior, each CP-table row still has a Dirichlet distribution but rows can now be dependent. The family includes "independent Dirichlet rows" as a special case. The DD priors do not form a conjugate family, and I do not have a tractable expression for the posterior distribution or the MP estimator. I instead derive the optimal linear (OL) estimator; i.e., $\hat{\theta}_{x|ab}$ is a weighted average of the sample proportions and prior means, with weights chosen to minimize the mean squared error MSE $= \sum_x E\{(\hat{\theta}_{x|ab} - \theta_{x|ab})^2\}$. The OL estimator can be viewed as an approximation to the MP estimator, since the latter minimizes the MSE among all estimators, not just linear estimators.

# 2 DEPENDENT DIRICHLET PRIORS

## 2.1 CONSTRUCTION

We can encode the joint distribution of a vector of discrete random variables $\boldsymbol{X} = \langle X_v \rangle_{v \in \mathcal{V}}$ as a belief net $\langle \mathcal{V}, \mathcal{A}, \boldsymbol{\Theta} \rangle$; i.e., a directed acyclic graph whose nodes $\mathcal{V}$ index the random variables and whose arcs $\mathcal{A}$ represent dependencies. My discussion focuses on a single child node $v \in \mathcal{V}$. I treat $v$ as fixed and suppress $v$ in the notation, writing $X = X_v$. Let $\mathcal{W} \subset \mathcal{V}$ be the set of immediate parents of node $v$, put $F_w = X_w$ for $w \in \mathcal{W}$, and let $\boldsymbol{F} = \langle F_w \rangle_{w \in \mathcal{W}}$ be the vector of parent variables. In a belief net, a variable $X$ is independent of its nondescendents, given its parents $\boldsymbol{F}$. The elements of the vector $\boldsymbol{\Theta}$ associated with child node $v$ are the CP-table entries

$$\theta_{x|\boldsymbol{f}} = Pr\{X = x \,|\, \boldsymbol{F} = \boldsymbol{f}, \boldsymbol{\Theta}\}.$$

Let $\mathcal{X}$ and $\mathcal{F} = \prod_{w \in \mathcal{W}} \mathcal{F}_w$ be the domains of $X$ and $\boldsymbol{F}$. I assume that the domains are finite. The CP-table for $X$ has $|\mathcal{X}|$ columns and $|\mathcal{F}| = \prod_{w \in \mathcal{W}} |\mathcal{F}_w|$ rows.

The CP-table entries are estimated using training data and (possibly) expert opinion. The latter information is incorporated using the Bayesian paradigm, where $\boldsymbol{\Theta}$ is modeled as a random variable and expert opinion is expressed through a prior distribution for $\boldsymbol{\Theta}$. CP-table rows are usually assumed to be independent, each with a Dirichlet distribution; i.e.,

$$\langle \theta_{x|\boldsymbol{f}} \rangle_{x \in \mathcal{X}} \sim \mathrm{Dir}(\alpha_{x|\boldsymbol{f}}, x \in \mathcal{X}). \tag{1}$$

The hyperparameters $\alpha_{x|\boldsymbol{f}}$ determine the nature and strength of the expert opinion. The prior means and covariances for the CP-table entries are:

$$E(\theta_{x|\boldsymbol{f}}) = \mu_{x|\boldsymbol{f}} = \frac{\alpha_{x|\boldsymbol{f}}}{\alpha_{\cdot|\boldsymbol{f}}}, \tag{2}$$

$$\mathrm{Cov}(\theta_{x|\boldsymbol{f}}, \theta_{y|\boldsymbol{f}}) = \frac{\mu_{x|\boldsymbol{f}}(\delta_{xy} - \mu_{y|\boldsymbol{f}})}{\alpha_{\cdot|\boldsymbol{f}} + 1}; \tag{3}$$

(Johnson and Kotz, 1972, page 233). Here and elsewhere, the dot notation represents summation over the subscript that the dot replaces ($\alpha_{\cdot|\boldsymbol{f}} = \sum_{x \in \mathcal{X}} \alpha_{x|\boldsymbol{f}}$) and $\delta$ is the Kronecker delta ($\delta_{xy} = 1$ if $x = y$ and $\delta_{xy} = 0$ otherwise). An absence of expert opinion is usually expressed by setting $\alpha_{x|\boldsymbol{f}} = 1$ for all $x \in \mathcal{X}$, which yields a uniform (flat) prior density. Stronger opinion is expressed through larger values of $\alpha_{\cdot|\boldsymbol{f}}$.

I now construct a larger family of prior distributions which permit dependencies among CP-table rows.

First some notation. Let $\eta$ have a Gamma($\alpha$) distribution if $\eta$ is a Gamma random variable with shape parameter $\alpha > 0$ and scale parameter 1; i.e., $\eta$ has density $\eta^{\alpha-1}\exp(-\eta)/\Gamma(\alpha)$ for $\eta > 0$. If $\alpha = 0$, then set $\eta = 0$. Consider a set of independent Gamma random variables $\eta_0(x)$, $\eta_w(x|f_w)$, and $\eta_2(x|\boldsymbol{f})$ with respective shape parameters $\alpha_0(x)$, $\alpha_w(x|f_w)$, and $\alpha_2(x|\boldsymbol{f})$, for $x \in \mathcal{X}, w \in \mathcal{W}$, and $\boldsymbol{f} = \langle f_w \rangle_{w \in \mathcal{W}} \in \mathcal{F}$. We thus have a set of $|\mathcal{X}|(1+\sum_{w\in\mathcal{W}} |\mathcal{F}_w|+|\mathcal{F}|)$ independent random variables. Define

$$\begin{aligned} \alpha_{x|\boldsymbol{f}} &= \alpha_0(x) + \sum_{w\in\mathcal{W}} \alpha_w(x|f_w) + \alpha_2(x|\boldsymbol{f}), \\ \eta_{x|\boldsymbol{f}} &= \eta_0(x) + \sum_{w\in\mathcal{W}} \eta_w(x|f_w) + \eta_2(x|\boldsymbol{f}). \end{aligned} \tag{4}$$

Note that $\eta_{x|\boldsymbol{f}} \sim$ Gamma($\alpha_{x|\boldsymbol{f}}$), with the variables $\eta_{x|\boldsymbol{f}}$ independent as $x$ varies with $\boldsymbol{f}$ fixed, but possibly dependent as $\boldsymbol{f}$ varies with $x$ fixed. A dependent Dirichlet (DD) prior assumes that

$$\theta_{x|\boldsymbol{f}} = \frac{\eta_{x|\boldsymbol{f}}}{\eta_{\cdot|\boldsymbol{f}}}. \tag{5}$$

Under this construction, each CP-table row has a Dirichlet distribution (Johnson and Kotz, 1972, page 231). Expressions (1) - (3) remain valid, but rows may now be dependent.

Two comments concerning the construction in (4). First, the Gamma variables are introduced solely to assist in constructing a joint distribution for $\langle\theta_{x|\boldsymbol{f}}\rangle$ and are otherwise not interpretable. Second, the notation in (4) mimics that of an additive statistical model. The term $\eta_0(x)$ can viewed as a "constant" term, common to all CP-table rows. The terms $\eta_w(x|f_w)$ represent the influence of individual parents. The term $\eta_2(x|\boldsymbol{f})$ can be viewed as an "error" or "residual" term, specific to each CP-table row.

In its general form, the DD family involves many hyperparameters; i.e., the $\alpha$ parameters. This level of generality may be useful for characterizing expert opinion in some situations, but a simpler subfamily will usually be adequate. Consider a *multiplicative* assumption defining the MDD family of priors:

$$\begin{aligned} \alpha_0(x) &= \alpha\mu_x\pi_0 \\ \alpha_w(x|f_w) &= \alpha\mu_x\pi_w \\ \alpha_2(x|\boldsymbol{f}) &= \alpha\mu_x\pi_2 \end{aligned} \tag{6}$$

with $\pi_\cdot = \mu_\cdot = 1$. The hyperparameters $\alpha$ and $\mu_x$ determine the marginal distribution for each CP-table row; i.e., $\alpha_{x|\boldsymbol{f}} = \alpha\mu_x$, $E(\theta_{x|\boldsymbol{f}}) = \mu_x$ and $\text{Var}(\theta_{x|\boldsymbol{f}}) = \mu_x(1-\mu_x)/(\alpha+1)$. A flat prior has $\alpha = |\mathcal{X}|$ and $\mu_x = 1/|\mathcal{X}|$ for all $x \in \mathcal{X}$. The hyperparameters $\pi_0, \pi_w, \pi_2$, and, to a lesser extent, $\alpha$ determine correlations between CP-table rows.

If there is no reason *a priori* to expect that $\theta_{x|\boldsymbol{f}}$ is related more strongly to one parent than to another, then it is reasonable to adopt a symmetric MDD prior:

$$\pi_w = \frac{\pi_1}{|\mathcal{W}|} \text{ for all } w \in \mathcal{W}, \quad \pi_0 + \pi_1 + \pi_2 = 1. \tag{7}$$

The MDD priors encompass several extremes:

- If $\pi_2 = 1$, then CP-table rows are independent.
- If $\pi_0 = 1$, then CP-table rows are all equal.
- If $\pi_2 = \pi_w = 0$ for all $w \neq w^*$, then $\theta_{x|\boldsymbol{f}} = \theta_{x|\boldsymbol{g}}$ when $f_{w^*} = g_{w^*}$. If, in addition, $\pi_0 = 0$, then $\theta_{x|\boldsymbol{f}}$ and $\theta_{x|\boldsymbol{g}}$ are independent when $f_{w^*} \neq g_{w^*}$.

## 2.2 COVARIANCES

To obtain the optimal linear estimator for $\theta_{x|\boldsymbol{f}}$, we need to calculate covariances $\sigma_{x\boldsymbol{fg}} = \text{Cov}(\theta_{x|\boldsymbol{f}}, \theta_{x|\boldsymbol{g}})$. The covariances can be expressed in terms of a function $\zeta$ defined as follows. Let $\eta_1, \eta_2, \eta_3$ be independent Gamma variables with respective shape parameters $\lambda_1, \lambda_2, \lambda_3$. Put $U = \eta_1/(\eta_1+\eta_2)$, $V = \eta_1/(\eta_1+\eta_3)$, and $\zeta(\lambda_1, \lambda_2, \lambda_3) = E(UV)$. The function $\zeta$ can be calculated by numerical integration. The following expression provides a convenient approximation for $\zeta$:

$$\frac{\lambda_1^*(\lambda_1^*+1)}{2\lambda_1^*+\lambda_2+\lambda_3}\left(\frac{1}{\lambda_1^*+\lambda_2+1} + \frac{1}{\lambda_1^*+\lambda_3+1}\right) \tag{8}$$

where

$$\lambda_1^* = \lambda_1\left(\frac{2\lambda_1+\lambda_2+\lambda_3}{2\lambda_1+\lambda_2+\lambda_3+1}\right).$$

The joint density of $(U, V)$ and the approximation formula are derived in a technical report (Hooper, 2004), which includes a derivation of the covariances for general DD priors. Results for MDD priors are as follows.

**Proposition 1.** Suppose we have an MDD prior. Put

$$\begin{aligned} \mathcal{W}_{\boldsymbol{fg}} &= \{w \in \mathcal{W} : f_w = g_w\}, \\ \gamma &= \pi_0 + \sum_{w\in\mathcal{W}_{\boldsymbol{fg}}} \pi_w + \delta_{\boldsymbol{fg}}\,\pi_2\ , \\ \rho(\alpha,\gamma) &= \frac{\alpha+1}{\alpha\gamma+1}\,\zeta(\alpha\gamma, \alpha(1-\gamma), \alpha(1-\gamma)). \end{aligned} \tag{9}$$

We then have

$$\text{Cov}(\theta_{x|\boldsymbol{f}}, \theta_{y|\boldsymbol{g}}) = \frac{\mu_x(\delta_{xy}-\mu_y)}{\alpha+1}\,\rho(\alpha,\gamma). \tag{10}$$

The function $\rho(\alpha,\gamma)$ equals the correlation between $\theta_{x|\boldsymbol{f}}$ and $\theta_{x|\boldsymbol{g}}$, which does not depend on $x \in \mathcal{X}$. This function is closely approximated by

$$\tilde{\rho} = \gamma - \left\{1 - 4(\gamma-0.5)^2\right\}\left\{0.5 - \rho(\alpha, 0.5)\right\}. \tag{11}$$

Table 1: Values needed to calculate expression (11) and the maximum approximation error.

| $\alpha$ | $0.5-\rho(\alpha,0.5)$ | $\max\lvert\tilde{\rho}-\rho\rvert$ |
|---|---|---|
| 2 | 0.071 | 0.007 |
| 3 | 0.054 | 0.005 |
| 4 | 0.044 | 0.004 |
| 5 | 0.037 | 0.003 |
| 10 | 0.021 | 0.002 |
| 20 | 0.011 | 0.0006 |

**Proof.** Expression (10) is derived in Hooper (2004). The quadratic approximation for the correlation is based on the following observations: $\gamma-\rho(\alpha,\gamma)\geq 0$ for all $\gamma$, equals 0 for $\gamma=0$ and 1, and attains its maximum near $\gamma=0.5$. Furthermore, $\rho(\alpha,\gamma)\to\gamma$ as $\alpha$ increases. Table 1 displays values of $0.5-\rho(\alpha,0.5)$ needed to calculate $\tilde{\rho}$, as well as the maximum error in the approximation. □

# 3 OPTIMAL LINEAR ESTIMATORS

Suppose we have a random sample of $n$ complete tuples $\langle X_v\rangle_{v\in\mathcal{V}}$. Let $m_{x\boldsymbol{f}}$ be the number of tuples with $(X,\boldsymbol{F})=(x,\boldsymbol{f})$. Put $n_{\boldsymbol{f}}=m_{\cdot\boldsymbol{f}}$, so $n=n_{\cdot}=m_{\cdot\cdot}$. The two random vectors $\langle n_{\boldsymbol{f}}\rangle$ and $\langle\theta_{x|\boldsymbol{f}}\rangle$ are assumed to be statistically independent. In the discussion that follows, the vector $\langle n_{\boldsymbol{f}}\rangle$ is treated as fixed; i.e., all distributions, expected values, variances, et cetera are implicitly assumed to be conditioned on $\langle n_{\boldsymbol{f}}\rangle$. Put $\mathcal{F}^a=\{\boldsymbol{f}\in\mathcal{F}:n_{\boldsymbol{f}}>0\}$. For $\boldsymbol{f}\in\mathcal{F}^a$, let $p_{x|\boldsymbol{f}}$ be the sample proportion $m_{x\boldsymbol{f}}/n_{\boldsymbol{f}}$. We then have

$$\begin{aligned} E(p_{x|\boldsymbol{f}}\,|\,\boldsymbol{\Theta}) &= \theta_{x|\boldsymbol{f}}, \\ \mathrm{Var}(p_{x|\boldsymbol{f}}\,|\,\boldsymbol{\Theta}) &= \theta_{x|\boldsymbol{f}}(1-\theta_{x|\boldsymbol{f}})/n_{\boldsymbol{f}}, \end{aligned}$$

and the variables $\{p_{x|\boldsymbol{f}}:\boldsymbol{f}\in\mathcal{F}^a\}$ are conditionally independent given $\boldsymbol{\Theta}$. We assume a prior distribution for $\langle\theta_{x|\boldsymbol{f}}\rangle$ with means $\mu_{x|\boldsymbol{f}}$ and covariances $\sigma_{x\boldsymbol{f}\boldsymbol{g}}$ but initially make no other assumptions about the prior.

Now fix $\boldsymbol{f}\in\mathcal{F}$, possibly with $n_{\boldsymbol{f}}=0$, and consider linear estimators for $\theta_{x|\boldsymbol{f}}$; i.e., weighted averages of sample proportions $p_{x|\boldsymbol{g}}$ and the prior means $\mu_{x|\boldsymbol{g}}$ for $\boldsymbol{g}\in\mathcal{F}^a$. As a notational device to simplify expressions, let the symbol $\boldsymbol{g}^*$ be an alternative label for the conditioning event $\boldsymbol{g}$, then put $\mathcal{F}^b=\{\boldsymbol{g}^*:\boldsymbol{g}\in\mathcal{F}^a\}$, $\mu_{x|\boldsymbol{g}^*}=\mu_{x|\boldsymbol{g}}$, and

$$p^*_{x|\boldsymbol{g}}=\begin{cases} p_{x|\boldsymbol{g}}-\mu_{x|\boldsymbol{g}}+\mu_{x|\boldsymbol{f}} & \text{if } \boldsymbol{g}\in\mathcal{F}^a \\ \mu_{x|\boldsymbol{g}} & \text{if } \boldsymbol{g}\in\mathcal{F}^b \end{cases}. \qquad (12)$$

This notation is motivated by the fact that $E\{p^*_{x|\boldsymbol{g}}\}=\mu_{x|\boldsymbol{f}}$ for all $\boldsymbol{g}\in\mathcal{F}^a$. Linear estimators can be expressed as

$$\hat{\theta}_{x|\boldsymbol{f}}=\sum_{\boldsymbol{g}\in\mathcal{F}^a\cup\mathcal{F}^b}a_{\boldsymbol{g}}p^*_{x|\boldsymbol{g}}, \qquad (13)$$

where the weights $a_{\boldsymbol{g}}$ are constants summing to one. This constraint $\sum a_{\boldsymbol{g}}=1$ is imposed because $\theta_{\cdot|\boldsymbol{f}}=1$ and $\hat{\theta}_{\cdot|\boldsymbol{f}}=\sum a_{\boldsymbol{g}}p^*_{\cdot|\boldsymbol{g}}=\sum a_{\boldsymbol{g}}$. Some of the weights may be negative; see Examples 4 and 5 below.

The following proposition gives an expression for a vector $\langle a_{\boldsymbol{g}}\rangle$ minimizing the mean squared error

$$\mathrm{MSE}=\sum_{x\in\mathcal{X}}E\left\{(\hat{\theta}_{x|\boldsymbol{f}}-\theta_{x|\boldsymbol{f}})^2\right\}. \qquad (14)$$

The proposition shows that it is always possible to choose an optimal vector $\langle a_{\boldsymbol{g}}\rangle$ with $a_{\boldsymbol{g}}=0$ for all $\boldsymbol{g}\in\mathcal{F}^b$ with $\boldsymbol{g}\neq\boldsymbol{f}^*$. Put $\mathcal{F}^c=\mathcal{F}^a\cup\{\boldsymbol{f}^*\}$, let $d=|\mathcal{F}^c|$, let $\boldsymbol{a}=\langle a_{\boldsymbol{g}}\rangle$ be a $d\times 1$ vector of weights, and let $\boldsymbol{1}_d$ be the $d\times 1$ vector of 1's. Let $\boldsymbol{B}$ be a $d\times d$ matrix with entries (for $\boldsymbol{g},\boldsymbol{h}\in\mathcal{F}^c$)

$$b_{\boldsymbol{g}\boldsymbol{h}}=\sum_{x\in\mathcal{X}}E\left\{(p^*_{x|\boldsymbol{g}}-\theta_{x|\boldsymbol{f}})(p^*_{x|\boldsymbol{h}}-\theta_{x|\boldsymbol{f}})\right\}. \qquad (15)$$

**Proposition 2.** Consider a restricted optimization problem: minimize the MSE among estimators

$$\hat{\theta}_{x|\boldsymbol{f}}=\sum_{\boldsymbol{g}\in\mathcal{F}^c}a_{\boldsymbol{g}}p^*_{x|\boldsymbol{g}}$$

with $\sum a_{\boldsymbol{g}}=1$. The MSE equals $\boldsymbol{a}'\boldsymbol{B}\boldsymbol{a}$. A vector $\boldsymbol{a}$ satisfying the constraint $\boldsymbol{a}'\boldsymbol{1}_d=1$ minimizes the MSE if and only if $\boldsymbol{B}\boldsymbol{a}=c\,\boldsymbol{1}_d$ for some $c\in\Re$. Solutions for this restricted problem are also optimal for the larger class of linear estimators (13). If $\boldsymbol{B}$ is nonsingular, then the solution is $\boldsymbol{a}=(\boldsymbol{1}_d'\boldsymbol{B}^{-1}\boldsymbol{1}_d)^{-1}\boldsymbol{B}^{-1}\boldsymbol{1}_d$ and the minimum MSE is $(\boldsymbol{1}_d'\boldsymbol{B}^{-1}\boldsymbol{1}_d)^{-1}$.

**Proof.** It is easily verified that

$$\boldsymbol{a}'\boldsymbol{B}\boldsymbol{a}=E\left\{\sum_x\left(\sum_{\boldsymbol{g}}a_{\boldsymbol{g}}(p^*_{x|\boldsymbol{g}}-\theta_{x|\boldsymbol{f}})\right)^2\right\}\geq 0$$

for all $\boldsymbol{a}\in\Re^d$, so $\boldsymbol{B}$ is a nonnegative definite symmetric matrix. If $\boldsymbol{a}'\boldsymbol{1}_d=1$, then MSE $=\boldsymbol{a}'\boldsymbol{B}\boldsymbol{a}$. Let $\boldsymbol{v}\in\Re^d$ be any vector satisfying $\boldsymbol{v}'\boldsymbol{1}_d=1$ and let $\boldsymbol{A}$ be any $d\times(d-1)$ matrix of full rank with $\boldsymbol{A}'\boldsymbol{1}_d=\boldsymbol{0}'$. Given $(\boldsymbol{v},\boldsymbol{A})$, the vector $\boldsymbol{a}$ satisfies the constraint if and only if $\boldsymbol{a}=\boldsymbol{v}+\boldsymbol{A}\boldsymbol{w}$ for some $\boldsymbol{w}\in\Re^{d-1}$. Differentiating $\boldsymbol{a}'\boldsymbol{B}\boldsymbol{a}$ with respect to $\boldsymbol{w}$, we see that $\boldsymbol{a}$ minimizes the MSE if and only if $\boldsymbol{A}'\boldsymbol{B}\boldsymbol{A}\boldsymbol{w}=-\boldsymbol{A}'\boldsymbol{B}\boldsymbol{v}$. Now the following statements are equivalent: $\boldsymbol{a}=\boldsymbol{v}$ minimizes the MSE $\iff\boldsymbol{w}=\boldsymbol{0}$ is a solution to the preceding equation $\iff\boldsymbol{A}'\boldsymbol{B}\boldsymbol{v}=\boldsymbol{0}\iff\boldsymbol{B}\boldsymbol{v}=c\,\boldsymbol{1}_d$. The solution when $\boldsymbol{B}$ is nonsingular is obvious.

It remains to show that any solution to the restricted problem minimizes the MSE over the larger family of estimators. Let $d_* = |\mathcal{F}^a \cup \mathcal{F}^b| = 2(d-1)$ and let $\boldsymbol{B}_*$ be the $d_* \times d_*$ matrix with entries $b_{\boldsymbol{gh}}$ for $\boldsymbol{g}, \boldsymbol{h} \in \mathcal{F}^a \cup \mathcal{F}^b$. We assume the elements in $\boldsymbol{B}_*$ are arranged so that

$$\boldsymbol{B}_* = \begin{pmatrix} \boldsymbol{B}_{11} & \boldsymbol{B}_{12} \\ \boldsymbol{B}_{21} & \boldsymbol{B}_{22} \end{pmatrix}$$

where $\boldsymbol{B}_{11}$ is the $d \times d$ matrix $\boldsymbol{B}$ defined previously. The discussion in the first part of the proof applies equally well to the unrestricted problem, so a vector $\boldsymbol{a}_* \in \Re^{d_*}$ satisfying the constraint $\boldsymbol{a}'_* \mathbf{1}_{d_*} = 1$ minimizes the MSE $\boldsymbol{a}'_* \boldsymbol{B}_* \boldsymbol{a}_*$ if and only if $\boldsymbol{B}_* \boldsymbol{a}_* = c\,\mathbf{1}_{d_*}$ for some $c \in \Re$. Now a solution $\boldsymbol{a}$ of the restricted problem has $\boldsymbol{B}_{11}\boldsymbol{a} = c\,\mathbf{1}_d$ for some $c \in \Re$. Furthermore, $\boldsymbol{B}_{21}\boldsymbol{a} = c\,\mathbf{1}_{d-2}$ for the same scalar $c$. This latter result follows a calculation, $b_{\boldsymbol{gh}} = \sigma_{\cdot\boldsymbol{ff}} - \sigma_{\cdot\boldsymbol{hf}}$ for $\boldsymbol{g} \in \mathcal{F}^b$ and $\boldsymbol{h} \in \mathcal{F}^a$, which shows that all of the rows in $\boldsymbol{B}_{21}$ equal the row in $\boldsymbol{B}_{11}$ corresponding to $\boldsymbol{g} = \boldsymbol{f}^*$. We thus have $\boldsymbol{B}_* \boldsymbol{a}_* = c\,\mathbf{1}_{d_*}$ for $\boldsymbol{a}'_* = (\boldsymbol{a}', \mathbf{0}')$. □

In practice, $\boldsymbol{B}$ is always nonsingular. In theory, we could have $\boldsymbol{B}$ singular and MSE $= 0$ in two situations. If $n_{\boldsymbol{f}} > 0$ and $\theta_{x|\boldsymbol{f}} \in \{0, 1\}$ with probability one, then $p_{x|\boldsymbol{f}} = \theta_{x|\boldsymbol{f}}$ with probability one, and $p_{x|\boldsymbol{f}}$ is the optimal estimator. If $\theta_{x|\boldsymbol{f}} = \mu_{x|\boldsymbol{f}}$ with probability one, then $\mu_{x|\boldsymbol{f}}$ is the optimal estimator. We could have $\boldsymbol{B}$ singular and MSE $> 0$ if $\mu_{x|\boldsymbol{g}} = 1$ for several values of $\boldsymbol{g} \neq \boldsymbol{f}$, say $\boldsymbol{g}_1$ and $\boldsymbol{g}_2$. We could then find a nonzero vector $\boldsymbol{c}$ satisfying $\boldsymbol{c}'\mathbf{1}_d = 0$ and $\sum c_{\boldsymbol{g}} p_{x|\boldsymbol{g}} = 0$ with probability one; e.g., take $\boldsymbol{c_g} = 1$ for $\boldsymbol{g} = \boldsymbol{g}_1$, $\boldsymbol{c_g} = -1$ for $\boldsymbol{g} = \boldsymbol{g}_2$, and $\boldsymbol{c_g} = 0$ otherwise. The optimal weights are not uniquely determined in this case since, if $\boldsymbol{a}$ is a vector minimizing the MSE subject to the constraint, then so is $\boldsymbol{a} + \boldsymbol{c}$.

It is straightforward to calculate $\boldsymbol{B}$ in terms of means and covariances. Recall that $\delta_{\boldsymbol{gh}}$ is the Kronecker delta and $\sigma_{\cdot\boldsymbol{gh}} = \sum_x \sigma_{x\boldsymbol{gh}}$. For $\boldsymbol{g}, \boldsymbol{h} \in \mathcal{F}^a$, we have

$$\begin{aligned} b_{\boldsymbol{gh}} &= (\delta_{\boldsymbol{gh}}/n_{\boldsymbol{g}}) \sum_{x \in \mathcal{X}} \{\mu_{x|\boldsymbol{g}}(1 - \mu_{x|\boldsymbol{g}}) - \sigma_{x\boldsymbol{gg}}\} \\ &\quad + \sigma_{\cdot\boldsymbol{ff}} + \sigma_{\cdot\boldsymbol{gh}} - \sigma_{\cdot\boldsymbol{gf}} - \sigma_{\cdot\boldsymbol{hf}}\ , \\ b_{\boldsymbol{gf}^*} &= \sigma_{\cdot\boldsymbol{ff}} + (\delta_{\boldsymbol{gf}^*} - 1)\sigma_{\cdot\boldsymbol{gf}}\ . \end{aligned} \tag{16}$$

Note that if $\boldsymbol{g} \in \mathcal{F}^c$ and $\boldsymbol{g} \neq \boldsymbol{f}$, then $b_{\boldsymbol{fg}} = 0$. These expressions simplify for MDD priors, where $\mu_{x|\boldsymbol{g}} = \mu_x$, $\sigma_{x\boldsymbol{gg}} = \sigma_{x\boldsymbol{ff}}$, and $p^*_{x|\boldsymbol{g}} = p_{x|\boldsymbol{g}}$ for $\boldsymbol{g} \in \mathcal{F}^a$. Let $\rho_{\boldsymbol{fg}} = \mathrm{Corr}(\theta_{x|\boldsymbol{f}}, \theta_{x|\boldsymbol{g}})$ given in (9). We have

$$\begin{aligned} b_{\boldsymbol{gh}} &= \sigma_{\cdot\boldsymbol{ff}}\{\delta_{\boldsymbol{gh}}\,\alpha/n_{\boldsymbol{g}} + 1 + \rho_{\boldsymbol{gh}} - \rho_{\boldsymbol{fg}} - \rho_{\boldsymbol{fh}}\}\ , \\ b_{\boldsymbol{gf}^*} &= \sigma_{\cdot\boldsymbol{ff}}\{1 + (\delta_{\boldsymbol{gf}^*} - 1)\rho_{\boldsymbol{fg}}\}\ . \end{aligned} \tag{17}$$

The common factor $\sigma_{\cdot\boldsymbol{ff}} = \sum_x \mu_x(1 - \mu_x)/(\alpha + 1)$ can be ignored when calculating the optimal weights $\langle a_{\boldsymbol{g}} \rangle$.

# 4 EXAMPLES AND DISCUSSION

This section illustrates Proposition 2 with five examples of MDD priors. In the first three examples, one can guess the optimal weight vector $\boldsymbol{a}$, then verify that $\boldsymbol{Ba} = c\,\mathbf{1}_d$.

**Example 1.** Suppose $\pi_2 = 1$, so CP-table rows are independent. We have $\rho_{\boldsymbol{gh}} = \delta_{\boldsymbol{gh}}$ for all $\boldsymbol{g}, \boldsymbol{h} \in \mathcal{F}^a$, $b_{\boldsymbol{gf}^*} = \sigma_{\cdot\boldsymbol{ff}}$ for all $\boldsymbol{g} \neq \boldsymbol{f}$, and $b_{\boldsymbol{ff}^*} = 0$. It is then easily verified that the optimal linear estimator is the usual mean posterior estimator

$$\hat{\theta}_{x|\boldsymbol{f}} = \frac{m_{x\boldsymbol{f}} + \alpha\mu_x}{n_{\boldsymbol{f}} + \alpha}.$$

**Example 2.** Suppose $\pi_0 = 1$, so all rows are equal and $\rho_{\boldsymbol{gh}} = 1$ for all $\boldsymbol{g}, \boldsymbol{h} \in \mathcal{F}^a$. In this case, $\boldsymbol{B}$ is a diagonal matrix and the optimal linear estimator is

$$\hat{\theta}_{x|\boldsymbol{f}} = \frac{1}{n + \alpha}\left\{\sum_{\boldsymbol{g} \in \mathcal{F}^a} n_{\boldsymbol{g}} p_{x|\boldsymbol{g}} + \alpha\mu_x\right\} = \frac{m_{x\cdot} + \alpha\mu_x}{n + \alpha}.$$

This solution yields the mean posterior estimator for the marginal probability $Pr\{X = x \,|\, \boldsymbol{\Theta}\}$, since each $\theta_{x|\boldsymbol{f}}$ is assumed to equal this marginal probability.

**Example 3.** Suppose $\pi_{w^*} = 1$ for some $w^* \in \mathcal{W}$, so $\theta_{x|\boldsymbol{f}} = \theta_{x|\boldsymbol{g}}$ if $f_{w^*} = g_{w^*}$ and rows are otherwise independent. The formula for the optimal linear estimator is similar to that in the previous example, but is now obtained by pooling counts over $\{\boldsymbol{g} \in \mathcal{F}^a : g_{w^*} = f_{w^*}\}$. This solution yields the mean posterior estimator for $Pr\{X = x \,|\, F_{w^*} = f_{w^*}, \boldsymbol{\Theta}\}$.

**Example 4.** This example explains why negative weights can occur. Suppose our child node has two binary parent nodes. Put $\boldsymbol{f} = \langle 0, 0 \rangle$, $\boldsymbol{g} = \langle 1, 0 \rangle$, $\boldsymbol{h} = \langle 1, 1 \rangle$, and $\boldsymbol{k} = \langle 0, 1 \rangle$. Suppose $n_{\boldsymbol{f}} = n_{\boldsymbol{g}} = n_{\boldsymbol{h}} = 10$ and $n_{\boldsymbol{k}} = 0$. Assume a symmetric MDD prior with $\pi_1 = 1$ and $\alpha = 2$. Recall the definition of $\gamma$ in expression (9). Here $\gamma_{\boldsymbol{fg}}$ equals the proportion of nodes where $\boldsymbol{f}$ and $\boldsymbol{g}$ agree, so $\gamma_{\boldsymbol{fg}} = \gamma_{\boldsymbol{gh}} = 0.5$ and $\gamma_{\boldsymbol{fh}} = 0$. From Table 1 we obtain $\rho_{\boldsymbol{fg}} = \rho_{\boldsymbol{gh}} = 0.500 - 0.071 = 0.429$ and $\rho_{\boldsymbol{fh}} = 0$. A matrix calculation yields optimal weights for $\hat{\theta}_{x|\boldsymbol{f}}$: $a_{\boldsymbol{f}} = 0.805$, $a_{\boldsymbol{g}} = 0.080$, $a_{\boldsymbol{h}} = -0.029$, and $a_{\boldsymbol{f}^*} = 0.144$. There is an intuitive explanation for the negative weight. If we observe $p_{x|\boldsymbol{g}} > p_{x|\boldsymbol{f}}$, then the positive correlation suggests that we should have $\hat{\theta}_{x|\boldsymbol{f}} > p_{x|\boldsymbol{f}}$. If we also observe $p_{x|\boldsymbol{h}} > p_{x|\boldsymbol{g}}$, then this weakens the previous conclusion since the larger value of $p_{x|\boldsymbol{g}}$ is partially explained by its correlation with a variable uncorrelated with $\theta_{x|\boldsymbol{f}}$.

Negative weights are potentially troublesome since one might obtain a probability estimate outside the interval $[0, 1]$. That cannot happen in Example 4, as long as $\mu_x$ is not too close to zero or one. While one cannot rule out the possibility of inadmissible probability estimates, I expect such estimates will occur rarely.

Table 2: Data and estimates.

| $\boldsymbol{f}$ | $n_{\boldsymbol{f}}$ | $m_{1\boldsymbol{f}}$ | $p_{1\|\boldsymbol{f}}$ | $\hat{\theta}^a_{1\|\boldsymbol{f}}$ | $\hat{\theta}^b_{1\|\boldsymbol{f}}$ |
|---|---|---|---|---|---|
| 000 | 22 | 12 | 0.545 | 0.542 | 0.561 |
| 001 | 5 | 2 | 0.400 | 0.429 | 0.487 |
| 010 | 15 | 9 | 0.600 | 0.588 | 0.594 |
| 011 | 8 | 4 | 0.500 | 0.500 | 0.510 |
| 100 | 14 | 14 | 1.000 | 0.937 | 0.939 |
| 101 | 9 | 8 | 0.889 | 0.818 | 0.830 |
| 110 | 5 | 4 | 0.800 | 0.714 | 0.777 |
| 111 | 0 | 0 | NaN | 0.500 | 0.701 |

Table 3: Weights ($\times 1000$) for $\hat{\theta}^b$.

| | 000 | 001 | 010 | 011 | 100 | 101 | 110 | 111 |
|---|---|---|---|---|---|---|---|---|
| 000 | 865 | 123 | 57 | 8 | 61 | 8 | –2 | –145 |
| 001 | 28 | 587 | 5 | 67 | 5 | 61 | –37 | 41 |
| 010 | 39 | 16 | 813 | 96 | 11 | –32 | 126 | 78 |
| 011 | 3 | 107 | 51 | 718 | –20 | 37 | 42 | 258 |
| 100 | 39 | 14 | 11 | –35 | 801 | 86 | 124 | 73 |
| 101 | 3 | 111 | –19 | 42 | 56 | 741 | 44 | 268 |
| 110 | –1 | –37 | 42 | 26 | 44 | 24 | 607 | 212 |
| 111 | 0 | 0 | 0 | 0 | 0 | 0 | 0 | 0 |
| $\mu$ | 23 | 80 | 40 | 79 | 42 | 73 | 97 | 216 |

A simple remedy is to move inadmissible estimates to zero or one, whichever is closer. The resulting adjusted estimator is actually a nonlinear estimator with MSE slightly smaller than that of the original linear estimator. A more complicated remedy would be to optimize weights subject to the constraint that all weights are nonnegative. I have not pursued this alternative since the resulting linear estimators would have larger MSE and hence be less efficient than the adjusted estimator.

**Example 5.** This last example illustrates calculations for a binary child node with three binary parents. We have a symmetric MDD prior with $\alpha = 2$ and $\mu_x = 0.5$. Estimates are given for two vectors $\boldsymbol{\pi} = \langle \pi_0, \pi_1, \pi_2 \rangle$: $\hat{\theta}^a$ for $\boldsymbol{\pi} = \langle 0, 0, 1 \rangle$ (independent rows), and $\hat{\theta}^b$ for $\boldsymbol{\pi} = \langle 0.25, 0.50, 0.25 \rangle$. The data and estimates are displayed in Table 2. The weights used to calculate $\hat{\theta}^b$ are displayed in Table 3. Note that the weights for the most dissimilar conditioning events are negative.

Larger counts $n_{\boldsymbol{g}}$ yield more information about parameters $\theta_{x|\boldsymbol{g}}$ correlated with $\theta_{x|\boldsymbol{f}}$, and hence should reduce the MSE for $\hat{\theta}_{x|\boldsymbol{f}}$. This relationship can be verified by examining the elements of $\boldsymbol{B}$. For each $\boldsymbol{g} \in \mathcal{F}^a$, the count $n_{\boldsymbol{g}}$ affects only the diagonal entry $b_{\boldsymbol{gg}}$, which is nonincreasing as $n_{\boldsymbol{g}}$ increases. It follows that $\boldsymbol{B}$ is nonincreasing (under the usual partial ordering for nonnegative definite symmetric matrices) as each $n_{\boldsymbol{g}}$ increases. Consequently, the minimum MSE is nonincreasing as a function of each $n_{\boldsymbol{g}}$.

The reduction in MSE from increasing $n_{\boldsymbol{g}}$ typically plateaus for $\boldsymbol{g} \neq \boldsymbol{f}$. As $n_{\boldsymbol{g}} \to \infty$, we obtain full information about $\theta_{x|\boldsymbol{g}}$, but this usually provides at best partial information about $\theta_{x|\boldsymbol{f}}$. For most prior distributions, the optimal linear estimator $\hat{\theta}_{x|\boldsymbol{f}}$ and the maximum likelihood estimator $p_{x|\boldsymbol{f}}$ are asymptotically equivalent as $n_{\boldsymbol{f}} \to \infty$, regardless of the limiting behaviour of the other $n_{\boldsymbol{g}}$. Here "most prior distributions" refers to those satisfying the condition that $\theta_{x|\boldsymbol{f}}$ cannot be expressed as a linear combination of its prior mean $\mu_{x|\boldsymbol{f}}$ and the values $\theta_{x|\boldsymbol{g}} - \mu_{x|\boldsymbol{g}} + \mu_{x|\boldsymbol{f}}$ for $\boldsymbol{g} \neq \boldsymbol{f}$. This condition is satisfied in Examples 1, 4, and 5, but not in Examples 2 and 3. The following is a more formal statement of the result for MDD priors.

**Proposition 3.** Consider an MDD prior with (i) $0 < \mu_x < 1$ and (ii) $\pi_2 > 0$ and/or $\pi_w > 0$ for all $w \in \mathcal{W}$. If $\langle a_{\boldsymbol{g}} \rangle$ is the optimal weight vector determined by the prior and the counts $\langle n_{\boldsymbol{g}} \rangle$, then $a_{\boldsymbol{f}} = 1 - \bigcirc(1/n_{\boldsymbol{f}})$ and $a_{\boldsymbol{g}} = \bigcirc(1/n_{\boldsymbol{f}})$ for all other $\boldsymbol{g} \in \mathcal{F}^c$, regardless of the limiting behaviour of $\langle n_{\boldsymbol{g}} \rangle_{\boldsymbol{g} \neq \boldsymbol{f}}$.

**Proof.** The fact that $b_{\boldsymbol{gf}} = 0$ for $\boldsymbol{g} \neq \boldsymbol{f}$ implies that the matrix $\boldsymbol{B}$ defined at (15) is block diagonal:

$$\boldsymbol{B} = \begin{bmatrix} b_{\boldsymbol{ff}} & \boldsymbol{0}' \\ \boldsymbol{0} & \boldsymbol{B}_1 \end{bmatrix}.$$

The term $b_{\boldsymbol{ff}}$ is of order $\bigcirc(1/n_{\boldsymbol{f}})$. The matrix $\boldsymbol{B}_1$ is positive definite symmetric, with smallest eigenvalue bounded away from zero for all $\langle n_{\boldsymbol{g}} \rangle$. Now the optimal weight vector $\boldsymbol{a}$ is proportional to $\boldsymbol{B}^{-1}\boldsymbol{1}_d$. The proof is completed by observing that $b_{\boldsymbol{ff}}^{-1}$ is of order $\bigcirc(n_{\boldsymbol{f}})$ and the largest eigenvalue of $\boldsymbol{B}_1^{-1}$ is uniformly bounded above for all $\langle n_{\boldsymbol{g}} \rangle$. $\square$

I developed optimal linear (OL) estimators for DD priors because traditional Bayesian estimators – the mean posterior (MP) and the maximum a posteriori (MAP) estimators – do not appear to be tractable. The MP estimator is defined as the conditional mean of $\theta_{x|\boldsymbol{f}}$ given the data. An equivalent definition is that the MP estimator minimizes the MSE among all estimators. One can thus regard the OL estimator as an approximation of the MP estimator restricted to the class of linear estimators. If the MP estimator is in fact a linear estimator, then it must agree with the OL estimator. This agreement holds for the MDD priors described in Examples 1 to 3, but the possibility of inadmissible probability estimates due to negative weights implies that agreement does not hold in general.

The MAP estimator is a vector maximizing the joint conditional density for $\langle \theta_{x|\boldsymbol{f}} \rangle$ given the data. This conditional distribution does not appear to have a tractable expression. In particular, consider an MDD

prior with $\pi_2 = 0$. The construction (4) shows that the vector $\langle \theta_{x|\boldsymbol{f}} \rangle$ is defined in terms of $|\mathcal{X}| \times \sum_{w \in \mathcal{W}} |\mathcal{F}_w|$ variables. The prior distribution for the parameter vector is thus constrained to a lower-dimensional non-linear manifold. The posterior distribution and MAP estimator would be constrained in the same manner. The OL and MP estimators are not constrained in this fashion.

# 5 SELECTING HYPERPARAMETERS

This section provides suggestions on how one might select DD prior hyperparameters in the absence of expert opinion. I would first focus on symmetric MDD priors, then specify values for $\alpha$, $\langle \mu_x \rangle$, and $\boldsymbol{\pi} = \langle \pi_0, \pi_1, \pi_2 \rangle$. It seems reasonable to choose a uniform Dirichlet distribution for each CP-table row; i.e., $\alpha = |\mathcal{X}|$ and $\mu_x = 1/|\mathcal{X}|$, so $\alpha\mu_x = 1$. It then remains to select $\boldsymbol{\pi}$ to reflect anticipated dependencies among rows. Consider two approaches to this problem.

## 5.1 A PRIORI COMPROMISE

The vector $\boldsymbol{\pi}$ lies in a simplex, with each $\pi_j \geq 0$ and $\sum \pi_j = 1$. One approach is to choose a vector somewhere in the middle of the simplex, in the hope that our selection will yield reasonably good results over a wide range of situations. This approach can be suppported with comparisions of MSE values conditioned on row counts $\langle n_{\boldsymbol{g}} \rangle$. Let $\boldsymbol{\pi}_{\text{select}}$ and $\boldsymbol{\pi}_{\text{true}}$ denote a selected vector and a hypothetically "true" vector. I suggest plotting values of the ratio of two mean squares: the MSE if $\boldsymbol{\pi}_{\text{select}}$ is used when $\boldsymbol{\pi}_{\text{true}}$ reflects the truth, and the minimum MSE achievable when $\boldsymbol{\pi}_{\text{true}}$ reflects the truth. This ratio is easily evaluated using Proposition 2. Let $\boldsymbol{B}_s$ and $\boldsymbol{B}_t$ denote the $\boldsymbol{B}$ matrices (17) corresponding to $\boldsymbol{\pi}_{\text{select}}$ and $\boldsymbol{\pi}_{\text{true}}$. Let $\boldsymbol{a} = (\mathbf{1}_d' \boldsymbol{B}_s^{-1} \mathbf{1}_d)^{-1} \boldsymbol{B}_s^{-1} \mathbf{1}_d$ be the optimal weight vector for $\boldsymbol{\pi}_{\text{select}}$. We then have

$$\text{MSE-ratio} = \frac{\text{MSE}}{\min \text{MSE}} = (\boldsymbol{a}' \boldsymbol{B}_t \boldsymbol{a})(\mathbf{1}_d' \boldsymbol{B}_t^{-1} \mathbf{1}_d). \quad (18)$$

The MSE-ratio represents the increase in MSE due to incorrect specification of the prior distribution. These values are illustrated in Figure 1, with plots for three $\boldsymbol{\pi}_{\text{select}}$ vectors. The points in each plot correspond to points $\boldsymbol{\pi}_{\text{true}}$ in the simplex with $\pi_j \in \{0.0, 0.1, 0.2, \ldots, 1.0\}$. It is assumed that there are four binary parents, hence $2^4 = 16$ CP-table rows, and that $n_{\boldsymbol{g}} = 3$ for each row. With such limited data, there is much to be gained by sharing information among rows if rows are in fact similar. The plot for $\boldsymbol{\pi}_{\text{select}} = \langle 0, 0, 1 \rangle$ shows how the usual estimator (assuming independent rows) performs comparatively well when rows appear to be independent, but poorly when rows tend to be similar. The plot for $\boldsymbol{\pi}_{\text{select}} = \langle 0, 1, 0 \rangle$ shows that the corresponding estimator does well when rows tend to be similar, but rather poorly when rows appear to be independent (with about twice the minimum MSE). The plot for $\boldsymbol{\pi}_{\text{select}} = \langle 0.25, 0.50, 0.25 \rangle$ suggests that the corresponding estimator performs comparatively well in both situations.

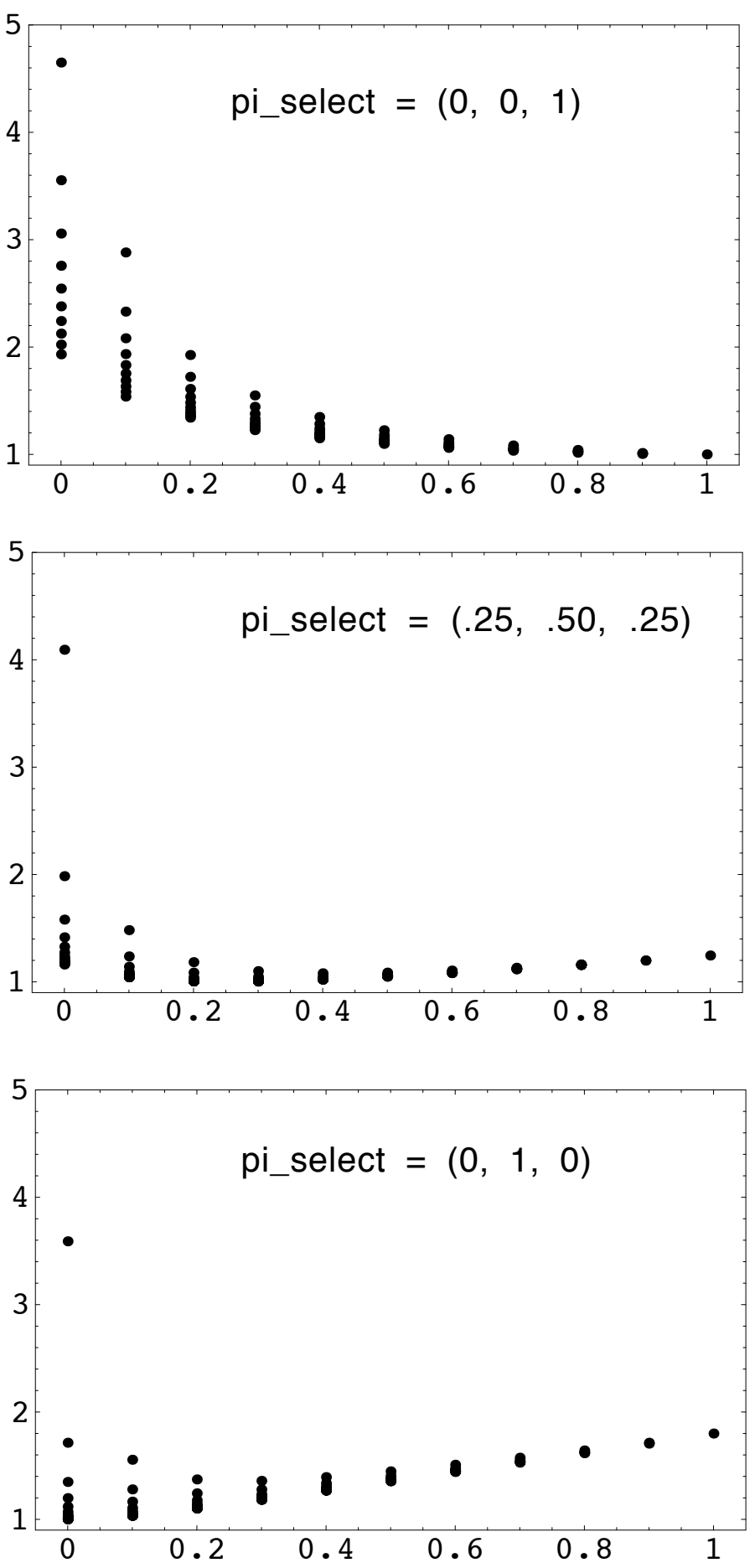

Figure 1: Plots of MSE-ratios for estimators determined by three selected priors. In all cases there are $2^4 = 16$ CP-table rows, each with $n_{\boldsymbol{g}} = 3$. The horizontal axis represents $\pi_2$, where $\boldsymbol{\pi}_{\text{true}} = \langle \pi_0, \pi_1, \pi_2 \rangle$. In each plot, when $\pi_2$ is fixed, the MSE-ratio increases as $\pi_0$ increases. There is one point that is out of the plot range in the plot for $\boldsymbol{\pi}_{\text{select}} = \langle 0, 0, 1 \rangle$: MSE-ratio $= 10.0$ when $\boldsymbol{\pi}_{\text{true}} = \langle 1, 0, 0 \rangle$.

## 5.2 EMPIRICAL BAYES

In an empirical Bayes analysis, hyperparameters defining a prior distribution are estimated from the data. We can estimate $\boldsymbol{\pi}$ using simple linear regression. Put $c_{\boldsymbol{fg}} = |\mathcal{W}_{\boldsymbol{fg}}|/|\mathcal{W}|$ = the proportion of nodes where $\boldsymbol{f}$ and $\boldsymbol{g}$ agree. From expressions (9) and (11), we have $\rho_{\boldsymbol{fg}} \approx \gamma_{\boldsymbol{fg}} = \pi_0 + \pi_1 c_{\boldsymbol{fg}}$ for $\boldsymbol{f} \neq \boldsymbol{g}$. Define

$$\hat{\rho}_{\boldsymbol{fg}} = \frac{\alpha + 1}{\mu_x(1-\mu_x)}(p_{x|\boldsymbol{f}} - \mu_x)(p_{x|\boldsymbol{g}} - \mu_x). \qquad (19)$$

It is easily verified that $E\{\hat{\rho}_{\boldsymbol{fg}}\} = \rho_{\boldsymbol{fg}}$ for $\boldsymbol{f} \neq \boldsymbol{g}$, both in $\mathcal{F}^a$. This identity suggests the following approach. Collect a set of paired values $(\hat{\rho}_{\boldsymbol{fg}}, c_{\boldsymbol{fg}})$ with $\boldsymbol{f} \neq \boldsymbol{g}$, fit a straight line, then set $\pi_0$ = the intercept, $\pi_1$ = the slope, and $\pi_2 = 1 - \pi_0 - \pi_1$.

The fitted regression line should be constrained so that $\pi_0 \geq 0$, $\pi_1 \geq 0$, and $\pi_0 + \pi_1 \leq 1$. This can be accomplished as follows. First fit an unconstrained regression line. If this satisfies the three constraints, then stop. Otherwise, fit three regression lines with various constraints: (i) $\pi_0 = 0$ and $0 \leq \pi_1 \leq 1$, (ii) $\pi_1 = 0$ and $0 \leq \pi_0 \leq 1$, and (iii) $\pi_0 + \pi_1 = 1$ and $0 \leq \pi_0 \leq 1$. Choose the fit that mimimizes the residual sum of squares.

I expect that the number of rows in a single CP-table will usually be too small to provide reliable estimates of correlations or $\boldsymbol{\pi}$. In particular, if there is just one parent node ($|\mathcal{W}| = 1$), then $c_{\boldsymbol{fg}}$ must be either 0 or 1, making the hyperparameter $\pi_1$ redundant. I suggest pooling the pairs $(\hat{\rho}_{\boldsymbol{fg}}, c_{\boldsymbol{fg}})$ from all CP-tables in the belief net, then fitting a single regression line. This approach yields a single choice of $\boldsymbol{\pi}$ to be used for all CP-tables.

Alternative methods of estimation may also be considered. One might treat $\boldsymbol{\pi}$ as a vector of tuning parameters, and select a value using cross-validation. The criterion could then depend on the particular application of the belief net; e.g., misclassification error.

# 6 RELATED METHODS

## 6.1 HIERARCHICAL PRIORS

If one is reluctant to choose a particular vector $\boldsymbol{\pi}$, an alternative approach is to quantify one's uncertainty by modeling $\boldsymbol{\pi}$ as a random vector. The resulting prior is then a mixture of MDD prior distributions. Priors constructed as mixtures over hyperparameters are called hierarchical priors. How does an additional layer of uncertainty affect the weights for optimal linear estimators? Consider modeling $\boldsymbol{\pi}$ as a random vector with mean $E(\boldsymbol{\pi})$. The entries $b_{\boldsymbol{gh}}$ in (17) are affected by replacing the correlations $\rho$ by their expected values. From (11) and (9), we see that $\rho$ is a convex function of $\gamma$, and $\gamma$ is a linear function of $\boldsymbol{\pi}$. Thus $E\{\rho(\alpha, \gamma)\} \geq \rho(\alpha, E\{\gamma\})$. It follows that weights $a_{\boldsymbol{g}}$ for $\boldsymbol{g} \neq \boldsymbol{f}$ will tend to be slightly larger than the weights for an MDD prior with $\boldsymbol{\pi} = E(\boldsymbol{\pi})$, since larger correlations imply more information to be shared among rows. The effect is generally small because the convexity in $\rho$ is slight; i.e., $\rho$ is roughly equal to $\gamma$. Consequently, it appears there is little to be gained by modeling $\boldsymbol{\pi}$ as a random vector.

Golinelli, Madigan, and Consonni (1999) proposed a family of prior distributions, called Hierarchical Partition Models (HPM), to share information among CP-table rows. Under an HPM, the set of rows $\mathcal{F}$ is randomly partitioned into subsets $\mathcal{F}_1, \ldots, \mathcal{F}_d$. Given a particular partition, the distribution of each row is a mixture of Dirichlet distributions, rows in different subsets are independent, and rows within the same subset are positively correlated; i.e., $\text{Corr}(\theta_{x|\boldsymbol{f}}, \theta_{x|\boldsymbol{g}}) > 0$ if $\boldsymbol{f}$ and $\boldsymbol{g}$ are both in $\mathcal{F}_k$. I have not carried out a quantitative comparison of HPM with MDD, but have two general comments relating the methods.

The first concerns the expression of prior opinion. Under the simplified version of HPM used in examples by Golinelli *et al.* (1999), all CP-table rows have the same distribution, all of the within-subset correlations are equal ($> 0.99$), and all partitions are equally likely. These assumptions imply that the prior distribution of $\langle\theta_{x|\boldsymbol{g}}\rangle_{\boldsymbol{g}\in\mathcal{F}}$ is exchangeable but not independent. The values $\theta_{x|\boldsymbol{g}}$ are expected to form an unknown number of tight clusters, with each pair $(\theta_{x|\boldsymbol{f}}, \theta_{x|\boldsymbol{g}})$ given an equal chance of belonging to the same cluster. The degree of similarity among conditioning events does not affect the prior probability that rows will be in the same cluster. In contrast, MDD assumes that rows with similar conditioning events will tend to be similar but not necessarily form clusters. In its more general form, HPM might exploit assumptions about "similarity of conditioning events" by assigning different correlations to different subsets, but it is not clear to me how this would be done.

The second comment concerns computational complexity. The number of partitions of $\mathcal{F}$ increases rapidly with $|\mathcal{F}|$, so HPM typically employs a Metropolis-Hastings algorithm to approximate the posterior means. Calculation of the optimal linear estimators under a DD prior is comparatively simple: first calculate the prior covariances (Proposition 1), then, for each CP-table row, calculate the $\boldsymbol{B}$ matrix and solve for the vector of weights (Proposition 2).

## 6.2 ALTERNATIVE MODELS

This paper considers prior distributions that lead to a sharing of information among sample proportions $p_{x|\boldsymbol{f}}$. The effect of the chosen prior is relatively weak, having a substantial effect on the estimator $\hat{\theta}_{x|\boldsymbol{f}}$ when $n_{\boldsymbol{f}}$ is small but little effect when $n_{\boldsymbol{f}}$ is large. Restrictions on the statistical model for belief net parameters impose

much stronger constraints on estimators, constraints that remain in full effect for all $\langle n_{\boldsymbol{f}} \rangle$. The following are three such models that are sometimes used. Assume here that $X$ and all of the parent variables comprising $\boldsymbol{F}$ are binary (0/1) variables. If $r = |\mathcal{W}|$ is the number of parents, then each of the following models reduces the number of nonredundant parameters from $2^r$ to $1 + r$.

- Logistic regression model: $\log\left(\theta_{1|\boldsymbol{f}}/\theta_{0|\boldsymbol{f}}\right) = \beta_0 + \boldsymbol{\beta}'\boldsymbol{f}$. Here $\beta_0$ and the entries in the vector $\boldsymbol{\beta}$ are unconstrained.
- Noisy-OR model: $\log\left(\theta_{1|\boldsymbol{f}}\right) = \beta_0 + \boldsymbol{\beta}'\boldsymbol{f}$. Here $\beta_0$ and each entry in $\boldsymbol{\beta}$ is $\leq 0$ to ensure that $\theta_{1|\boldsymbol{f}} \leq \theta_{1|0\ldots0} \leq 1$. Sometimes $\beta_0$ is set to zero, so $\theta_{1|0\ldots0} = 1$.
- Decision tree model: Suppose $\boldsymbol{F} = \langle A, B, C \rangle$. A decision tree model might assert that (i) if $A = 1$, then $B$ and $C$ are irrelevant, and (ii) if $A = 0$ and $B = 1$, then $C$ is irrelevant. Thus $\theta_{x|1bc} = \theta_{x|111}$ and $\theta_{x|01c} = \theta_{x|011}$. Decision tree models are often deterministic; i.e., each $\theta_{x|\boldsymbol{f}}$ is either zero or one.

The list above does not include log-linear models for contingency tables (Bishop, Fienberg, and Holland, 1975, Chapter 2), which express conditional independence assumptions. Assumptions of this kind are reflected in the structure of the belief net.

# 7 CONCLUSION

This paper presents alternatives to the widely used local independence assumption. The new prior distributions are motivated by the notion that rows representing similar conditioning events tend to have similar conditional probabilities. This general idea is employed in other statistical applications; e.g., in the analysis of sample survey data, estimates of parameters for "small areas" are often improved by "borrowing strength" from similar areas (Rao, 2003).

DD priors quantify *a priori* opinion about similarities among CP-table rows. Two methods are given to assist in selecting a particular prior from the DD family. Optimal linear estimators provide an effective method for combining prior opinion with empirical data.

Plans for further work include evaluation of the proposed methodology in applications of belief nets. DD priors were originally motivated by the problem of estimating belief net queries. CP-table parameters are typically not of direct interest, but are used to compute more relevant conditional probabilities of the form $Pr\{\boldsymbol{H} = \boldsymbol{h} \,|\, \boldsymbol{E} = \boldsymbol{e}\}$, where the "hypothesis" $\boldsymbol{H}$ and the "evidence" $\boldsymbol{E}$ each involve one or more belief net variables. These "query" probabilities can be expressed in terms of CP-table parameters, and are usually estimated by plugging in parameter estimates. I expect DD priors will tend to improve query estimates by exploiting a bias-variance trade-off.

Van Allen, Greiner, and Hooper (2001) developed approximate Bayesian credible intervals (error-bars) for belief net queries. A topic under current investigation concerns the extension of their method (which assumes independent Dirichlet rows) to DD priors. This extension involves the approximation of covariances for the posterior distribution of CP-table parameters.

## Acknowledgements

This work was partially supported by the Natural Sciences and Engineering Research Council of Canada. I am grateful to the referees, whose comments led to several improvements.

## References

Y. Bishop, S. Fienberg, and P. Holland (1975). *Discrete Multivariate Analysis: Theory and Practice.* Cambridge, MIT Press.

D. Golinelli, D. Madigan, and G. Consonni (1999). Relaxing the local independence assumption for quantitative learning in acyclic directed graphical models through hierarchical partition models. In *Proceedings of Uncertainty-99, The Seventh International Workshop on Artificial Intelligence and Statistics*, 203-208.

D. Heckerman (1999). A tutorial on learning with Bayesian networks. In *Learning in Graphical Models*, M. Jordan, ed.. Cambridge, MIT Press.

P. M. Hooper (2004). Dependent Dirichlet priors and optimal linear estimators for belief net parameters. Technical report, University of Alberta.

N. L. Johnson and S. Kotz (1972). *Distributions in Statistics: Continuous Multivariate Distributions.* New York, John Wiley & Sons.

J. N. K. Rao (2003). *Small Area Estimation (Methods and Applications).* New York, John Wiley & Sons.

D. J. Spiegelhalter and S. L. Lauritzen (1990). Sequential updating of conditional probabilities on directed graphical structures. *Networks* **20**: 579-605.

T. Van Allen, R. Greiner, and P. M. Hooper (2001). Bayesian error-bars for belief net inference. In *Proceedings of the Seventeenth Conference on Uncertainty in Artificial Intelligence (UAI-01)*, 522-529.